\documentclass{ws-ijmpd}
\usepackage[utf8x]{inputenc}
\usepackage[english]{babel}
\usepackage{graphicx}
\usepackage{float}
\usepackage[super,compress]{cite}
\usepackage{setspace, longtable, amstext}
\usepackage{selinput}

\begin{document}

\markboth{Authors' Names}
{Instructions for Typing Manuscripts (Paper's Title)}

\catchline{}{}{}{}{}
%
%%%%%%%%%%%%%%%%%%%%%%%%%%%%%%%%%%%%%%%%%%%%%%%%%%%%%%%%%%%%%%%%%%%%

\title{On the structure of galactic halos and the microwave temperature maps}  

\author{A.Amekhyan$^1$\footnote{amekhyan@yerphi.am}}

\address{Center for Cosmology and Astrophysics, Alikhanian National Laboratory, Yerevan, Armenia
}

\maketitle

\begin{history}
\received{16 March 2019}
\accepted{25 March 2019}
\end{history}

\begin{abstract}
The study of Planck microwave temperature maps towards several nearby spiral edge-on galaxies had revealed frequency independent temperature asymmetry of Doppler origin in their halos. On an example of M31, as of relatively detailed studied galaxy, the contribution of the dust component in that effect is studied using the models of dust emission and the phenomenological profiles of the dark matter configurations. The results are in accordance with those obtained from the microwave temperature asymmetry data, thus indicating the possible contribution of dust among other radiation mechanisms in revealing the dark halo parameters.    
\end{abstract}


\ccode{PACS numbers: 98.62.Gq}

%\tableofcontents

\section{Introduction}

The importance of studies of galactic halos is essentially determined by their key role in the deciphering of the nature of dark matter. The physical parameters of the halos not only determine the dynamics of galaxies including of the disks in spiral galaxies \cite{Kr,Feld}, but also provide informative criteria to constrain the models of evolution of the cosmological perturbations \cite{Ok}.  

Among the recent studies addressing the structure of galactic halos are those based on the analysis of the microwave temperature asymmetry of nearby galaxies using several frequency bands of the Planck data \cite{DP14,DP15,DP16,G15,G18}. Namely, first for M31, then for several other Local Supercluster spiral edge-on galaxies it was shown that, the temperature asymmetry both in the disk and halo extends up to 130kpc.
 While the essentially frequency independent temperature asymmetry is indicating the Doppler nature of the effect, the Doppler induced anomaly itself can be due to different emission mechanisms of interstellar medium (ISM). One of the main components of ISM is the interstellar thermal dust, which emits especially at high frequencies. The possible contribution of the thermal dust in that effect is studied below using high frequency data of Planck. At the same time, dark matter configuration models have been actively developed, among those are Navarro-Frenk-White (NFW) \cite{N}, Moore \cite{M}, Burkert \cite{B} models, so we will deal with those models in our analysis.  As for M31 dark halo profile, it was studied e.g. in \cite{Tam,Sof,TT}.

The present study concentrates on M31 galaxy, to probe the physical effects which can contribute to the microwave temperature asymmetry. Various models have been proposed for interstellar dust, which fix the properties of dust particles i.e. the chemical composition, dust grain size, distribution, etc. For our analysis we use the silicate-graphite-PAH model proposed in \cite{DL} (hereafter DL07), as well as Planck’s dust emission maps. According to Planck team \cite{P1,P2,P3}, these models are described by single modified blackbody spectrum (MBB), from which we obtain radial variation of dust emission optical depth, using Planck’s dust emission map at three frequencies.

The paper is organized as follows. Section 2 describes the dust models and MBB spectrum. Planck maps are described in Section 3, which then are used in calculations. Section 4 is focused on the optical depth values derived using MBB spectrum. In section 5 the dark matter distribution profiles are analyzed. CMB temperature asymmetry values as well as the rotation velocities obtained via Doppler effect induced asymmetry formula are given. The conclusions are summarized in section 6.
 
\section{Interstellar thermal dust emission}

\subsection{The DL07 model}

In this model, the dust is assumed to consist of a mixture of carbonaceous grains and amorphous silicate grains. In interstellar medium dust is heated from starlight and re-emits it in infrared or far-infrared wavelengths. DL07 model \cite{DL} has several parameters  which completely define all the properties of dust grains. There are two types of dust grains: small grains ($a\leq300Å$)  and large grains ($a\approx2200Å$). Small grains are warmer than big grains. According to \cite{Hil} major part of the emission comes from large grains, which is dominating part of all interstellar dust. DL07 model has several parameters, among those are are the dust mass surface density $\Sigma_{M_d}$, dust optical extinction $A_{\nu}$, starlight intensity  $U_{min}$, the dust mass fraction in small PAH (Polycyclic Aromatic Hydrocarbon particles) grains $q_{PAH}$. Here the quantity $q_{PAH}$ is defined as PAH’s fraction of the total dust mass. PAHs are emitted at 3.3$\mu$m, 6.2$\mu$m, 7.7$\mu$m, 8.6$\mu$m, and 11.3$\mu$m wavelengths and we are not interested in their contribution, since we consider only three wavelengths (350$\mu$m, 550$\mu$m, 850$\mu$m). However, the exact modeling of PAHs is important for definition of total dust mass \cite{DL}.

Important parameter is the dust Spectral Energy Distribution (SED) shape as compared with the internal mechanism from which SED emerges, since we need to calculate the radial variation of the dust emission optical depth.

Therefore, for our analysis we use only the dust mass surface density (or equivalent flux density from maps) to determine the dust velocity dispersion. As suggested by \cite{Hil} there is relationship between dust mass and optical depth
\begin{equation}
M_{dust}=\frac{\tau D^{2}\Omega}{k},
\end{equation}
where $\tau$ is the optical depth, $k$ is the dust mass absorption coefficient, which depends not only on given frequency, but also on chemical composition and structure of the dust grains; for example, at 353 GHz (850$\mu m$) $K_{\nu}=0.43 \pm 0.04 cm^{2}g^{-1}$ \cite{Abergel}, $D$ is the distance from the dust grains to observer, $\Omega$ is the solid angle. Optical depth ($\tau$) shows intensity reduction of light emitted by dust grains and propagating over line-of-sight. The values of dust mass surface density, as well as of other parameters for DL07 model are publicly available in the Planck Legacy Archive and the parameters' processing is described in details in \cite{P5}. In Table 1 we present the values of dust mass surface density for M31 obtained based on the Planck maps.

\subsection{The modified blackbody spectrum}

As mentioned in the several Planck papers \cite{P1,P2,P3} the DL07  and GNILC-derived SEDs are compatible with MBB spectrum, which is obtained empirically from Planck observations of dust. For optically thin sources (like M31) the MBB function is as follows 
\begin{equation}
S_{dust}(\nu)=2h\frac{\nu^3}{c^2}\frac{1}{e^{\frac{h\nu}{kT_d}}-1}\tau (\frac{\nu}{\nu_0})^{\beta},
\end{equation}
where $S_{dust}(\nu)$ is the dust emission flux density, $T_{d}$ is the dust temperature and $\beta$ is the spectral index. Through spectral index we can identify the dust model. For  DL07 model spectral index value of 2 for any radii is assumed. The dust temperature  changes in more complex way. The Planck data indicate that, there is an anti-correlation between $T_{d}$ and $\beta$. $T_{d}$ decreases, while  $\beta$ increases at large galactocentric distances.  This anti-correlation has been examined in several studies \cite{Dupac,Aber,Abergel}.The origin of anti-correlation is non-physical and is caused by instrumental noise or uncertainties at MBB spectrum fitting \cite {Juvela,Shetty}. However, as reported in \cite{Ade}, the spectral index for millimeter wavelengths is insensitive to temperature, while for FIR wavelengths the degeneracy is significant due to data uncertainties. The values of the dust temperature and spectral index  are given in Table 1. On the other hand, the temperature may change even for fixed value of spectral index. Therefore, we calculate the optical depth for two different cases:
\begin{enumerate}
\item Both $T_{d}$ and $\beta$ are constant at 19.635 K. Keeping fixed $\beta$, we vary the temperature. The data on temperature for different galactocentric distances are taken from Planck GNILC temperature maps.
\item Using GNILC maps for  $T_{d}$ and $\beta$  we calculate the radial variation of the optical depth vs temperature and spectral index. We also cover optical depth, when  $\beta=2$, $T_{d}\neq$ const. These maps are available in Planck Legacy Archive, their processing is described in \cite{Agh}.
 \end{enumerate}
 
\begin{table}[h!]    %%%1
\tbl {Dust temperature, spectral index and mass surface density.}
{
\centering
\begin{tabular}{ |p{2cm}||p{2cm}| |p{2cm}||p{2cm}| }
\hline
  Distance\newline r (kpc)
   &Temperature  $T_{d}(K)$
   &Spectral index $\beta$
   &Mass surface density $\frac{M_{\odot}}{kpc^2}(10^{4})$\\
\hline
   40&19.768&1.516&8.10 \\
   60&19.600&1.533&7.358\\
   80&19.541&1.541&7.208\\
  100&19.517&1.545&7.188\\
  \hline
  \end{tabular}
}	
\end{table}

 \section{Planck maps}

During the calculation of the optical depth we use several maps produced by Planck. These maps are publicly available in Planck Legacy Archive\footnote{http://pla.esac.esa.int}. All maps, which we use during this study are given in Hierarchical Equal Area iso-Latitude Pixelization (HEALPix) with galactic coordinate system \cite{Gor}. All maps have $N_{side}$=2048 resolution, so the maps have $12 \times 2048 \times 2048= 50331648$ pixels. For our analysis we extracted 2D projected maps centered on M31 ( $l =121.17^{0}$, $b = −21.57^{0}$), with $10^{0}$ width region.

\textit{DL07 and GNILC maps: } The Generalized Needlet Internal Linear Combination (GNILC) method uses spatial information to separate dust emission and cosmic infrared background (CIB). The component-separation technique details are discussed in \cite{Agh}. It is important to note, that GNILC maps are based on prior assumptions on the other foreground emissions (CMB, CIB, free-free, etc). In contrast, DL07 model-predicted fluxes are based on an individual dust model. Despite this, for galactic halo the MBB well describes both DL07 and GNILC methods. GNILC maps also have offsets at each frequencies(0.556 MJy/sr for 857 GHz, 0.335 MJy/sr for 545 GHz and 0.124 MJy/sr for 353 GHz), which we have taken into account. Besides the three frequency maps we also use GNILC spectral index and temperature maps, which have been obtained using MBB fitting procedure. For flux density $S_{dust}(\nu)$ values we use both DL07 and GNILC maps in three bandwidths.
%\begin{figure}[t!]
%\centering
%\includegraphics[width=\linewidth]{m31.eps}
%	\label{fig:m31}
%\caption{\footnotesize 857GHz Planck thermal dust map of m31}
%\end{figure}

In our calculations we started from 35 kpc distance from M31 center and substituted each pixel with the average value of the equidistant pixels.

\section{ Dust emission optical depth }

As mentioned above, we use MBB spectrum for obtaining the spatial variations of optical depth in M31 halo. We begin with DL07 model, where dust spectral index has fixed value of 2. Dust grain in thermal equilibrium with radiation field has uniform temperature distribution with average 19.635K (so here we consider this temperature value for M31 halo in the range of 30-100 kpc). Then we assume varying temperature (from GNILC map). The result of our calculations are given in the Table 2. The table shows, that the optical depth gradually decreases with increasing radius.  As we can see from Table 1, the temperature also decreases within large radii, but it does not affect the value of optical depth. Namely, in this case optical depth depends only on the dust emission flux density. 

\begin{table}[h!]   %%% 2
\tbl{The optical depth for DL07 model.}
{
\centering
\begin{tabular}{ |p{2cm}||p{2cm}| |p{2cm}||p{2cm}| }
 \hline
  \multicolumn{4}{|c|}{DL07 $\beta=2,$ $T=19.635K$} \\
 \hline
 Distance\newline r (kpc)
 
 &Optical depth\newline$\tau_{857}$ 
 
 &Optical depth\newline$\tau_{545}$
 
 &Optical depth\newline$\tau_{353}$ \\
 
  \hline
  40&$7.671\times10^{-4}$&$1.616\times10^{-3}$&$2.629\times10^{-3}$\\
 
  60&$6.122\times10^{-4}$&$8.345\times10^{-4}$&$1.465\times10^{-3}$\\
 
  80&$4.166\times10^{-4}$&$4.314\times10^{-4}$&$1.066\times10^{-3}$\\
 
  100&$3.456\times10^{-4}$&$1.911\times10^{-4}$&$1.508\times10^{-4}$\\
  \hline
  \hline
\multicolumn{4}{|c|}{DL07  $\beta=2,$ $T\neq const$}\\ 
   \hline
Distance\newline r (kpc)

&Optical depth\newline $\tau_{857}$     
 
&Optical depth\newline $\tau_{545}$

&Optical depth\newline $\tau_{353}$\\

  \hline

   40&$7.620\times10^{-4}$&$1.605\times10^{-3}$&$2.612\times10^{-3}$\\
 
   60&$6.133\times10^{-4}$&$8.366\times10^{-4}$&$1.468\times10^{-3}$\\
  
	 80&$4.187\times10^{-4}$&$4.337\times10^{-4}$&$1.071\times10^{-3}$\\
 
  100&$3.480\times10^{-4}$&$1.923\times10^{-4}$&$1.517\times10^{-4}$\\
	\hline
  \end{tabular} \label{tab:2}}
	
\end{table}

Next, we calculate optical depth using GNILC maps. In this case both spectral index and temperature change. There is anti-correlation between $T_{d}$ and $\beta$ reported by Planck collaboration \cite{P1,P2}. The spectral index of dust increases with radius, which means, that dust emission properties depend on the temperature, and the heating rate of dust depends on the spectral index. The corresponding values of the optical depth in this case are given in Table 3 and Table 4. As in DL07, in this case also there is almost no temperature dependency of optical depth. Optical depth is only determined through spectral index and flux density (derived by GNILC method). 

\begin{table}[h!]   %%% 3
\tbl{ The optical depth for GNILC  model.}
{
\centering
\begin{tabular}{ |p{2.5cm}||p{2.7cm}| |p{2.5cm}||p{2.5cm}| }
 \hline
  \multicolumn{4}{|c|}{GNILC   $\beta=2$, $T\neq const$} \\
  \hline
 Distance\newline r[kpc]
 &Optical depth\newline$\tau_{857}$
 &Optical depth\newline$\tau_{545}$
 &Optical depth\newline$\tau_{353}$\\
  \hline
    
   40&$7.999\times10^{-4}$&$1.482\times10^{-3}$&$2.214\times10^{-3}$\\

   60&$6.421\times10^{-4}$&$8.630\times10^{-4}$&$1.208\times10^{-3}$\\
 
 80&$4.786\times10^{-4}$&$3.177\times10^{-4}$&$4.003\times10^{-4}$\\
 
 100&$3.587\times10^{-4}$&$1.962\times10^{-4}$&$1.083\times10^{-4}$\\
  \hline
  \end{tabular} \label{table:3}}
 \end{table}

\begin{table}[h!]    %%% 4
\tbl{The optical depth for GNILC  model.}
{
\centering 
\begin{tabular}{ |p{2.5cm}||p{2.5cm}||p{2.5cm}||p{2.5cm}|  }
\hline
\multicolumn{4}{|c|}{GNILC   $\beta\neq const$, $T\neq const$} \\
\hline
Distance\newline r (kpc)
&Optical depth\newline $\tau_{857}$
&Optical depth\newline $\tau_{545}$ 
&Optical depth\newline $\tau_{353}$\\
\hline
  40&$2.100\times10^{-2}$&$3.152\times10^{-2}$&$3.771\times10^{-2}$\\
  60&$1.503\times10^{-2}$&$1.648\times10^{-2}$&$1.862\times10^{-2}$\\
  80&$1.084\times10^{-2}$&$5.778\times10^{-3}$&$5.989\times10^{-3}$\\
 100&$7.850\times10^{-3}$&$3.493\times10^{-3}$&$1.564\times10^{-3}$\\
\hline
\end{tabular} \label{table:4}}
\end{table}

\section{Dark matter distribution profiles for M31}

M31 is one of the most studied galaxy in the Local Group. Various authors have suggested different dark matter distribution profiles for M31 halo. Here, in this paper we use the Navarro-Frenk-White (NFW) \cite{N}, Moore \cite{M} and Burkert \cite{B} models. The corresponding density and velocity distributions are presented below:

\begin{equation}
\rho _{NFW} (r)=  \frac{\rho_c}{(\frac{r}{r_c})\Big(1+\big(\frac{r}{r_c}\big)\Big)^2},
\end{equation}

\begin{equation}
V_{NFW}^2(r)=4\pi G\rho_c \frac{r_c^3}{r}\big[ln\big(1+\frac{r}{r_c}\big)-
\frac{\frac{r}{r_c}}{1+\frac{r}{r_c}}\big],
\end{equation}

\begin{equation}
\rho_{Moore}(r)=\frac{\rho_c}{\big(\frac{r}{r_c}\big)^{1.5}\Big[1+\big(\frac{r}{r_c}\big)^{1.5}\Big]},
\end{equation}

\begin{equation}
V_{Moore}^2(r)=\frac{8}{3}\pi G \rho_c\frac{r_c^3}{r} ln\Big[1+\big(\frac{r}{r_c}\big)^{1.5}\Big],
\end{equation}

\begin{equation}
\rho_{Burkert}(r)= \frac{\rho_c}{\Big(1+\frac{r}{r_c}\Big)\Big[1+\big(\frac{r}{r_c}\big)^2\Big]},
\end{equation}

\begin{equation}
V_{Burkert}^2(r)=2\pi G\rho_c \frac{r_c^3}{r}  \Big\{\Big[ln\big(1+\frac{r}{r_c}\big)  \sqrt{1+\big(\frac{r}{r_c}\big)^2}    \Big] -\arctan(\frac{r}{r_c}) \Big\},
\end{equation}
where  $r_c=\frac{r_{vir}}{r}$ and $\rho_{c}$ are characteristic radius and density. 
The virial radius $r_{vir}$ is defined as the radius at which the mean density is equal to the overdensity constant  ( $\Delta_{vir}$ ) multiplied by the critical density of the universe

Corresponding virial  mass is defined as 
\begin{equation}
M_{vir}=\frac{4\pi}{3}\Delta_{vir}\rho_{crit}R_{vir}^3.
\end{equation}
Here $\rho_{c}$ and $r_{c}$ are free parameters, fixed by fitting procedure. In Table 5 the corresponding fitting values for these parameters are suggested by \cite{TT}.

\vspace{0.5cm}
\begin{table}[h!]   %%% 5
\tbl{Best fit values of the DM profiles of \cite{TT}.}
{\centering
\begin{tabular}{ |p{1.5cm}| |p{2.5cm}| |p{1.5cm}| |p{2.5cm}| |p{1.5cm}| }
 \hline
 Profile& $\rho_{c} (M_{\odot}(pc)^{-3})$&$r_{c}(kpc)$&$M_{vir}(10^{11}M_{\odot})$&$R_{vir}(kpc)$ \\
\hline
 
 NFW&$1.74*10^{-2}$&12.5&6.93&146.5\\
 
 Moore&$2.05*10^{-3}$&25.0&7.38&149.6\\
 
 Burkert&$5.72*10^{-2}$&6.86&5.16&132.8\\
\hline
\end{tabular} \label{table:5}}
%\caption{\footnotesize  Best fit values of the DM profiles by [22]} \label{table:5}
\end{table}

On the other hand,   we calculate the M31 halo rotational velocities using the Doppler effect induced temperature asymmetry formula for optically thin halo \cite{DP95} 
\begin{equation}
\frac{|\Delta T|}{T}=\frac{2vsini}{c}\tau.
\end{equation} 
 Here the galaxy inclination angle is $i=77^{0}$ and $\tau$ is the optical depth, which depends on the  dust parameters and frequencies. $T$ is the dust temperature, which changes at large galactocentric distances (see Table 1), and finally $|\Delta T|$ is the CMB temperature asymmetry in $\mu$K\footnote{Actually, the Planck maps for 857 GHz and 545 GHz frequencies are given in  MJy/sr units, which we have converted to $\mu$K via Planck High Frequency Instrument (HFI) unit conversion and colour correction coefficients \cite{P4}.}.

\begin{table}[h!]   %%% 6
\tbl{CMB asymmetry data by Planck\cite{DP16}.}
{
\centering
\begin{tabular}{ |p{2cm}||p{2.2cm}||p{2cm}||p{2cm}|}
 \hline
 Distance\newline r (kpc)
&$\nu=857GHz$\newline$\frac{\Delta T}{T}$($\mu K$)$(10^{-2}$)
&$\nu=545GHz$\newline$\frac{\Delta T}{T}$($\mu K$)
&$\nu=353GHz$\newline$\frac{\Delta T}{T}$($\mu K$)\\ 
\hline
40&4.68&0.378&6.11\\

60&5.0&0.378&6.09  \\

80&3.51&0.277&3.27 \\

100&2.12&0.176&0.79 \\
\hline
\end{tabular} \label{table:6}}
\end{table}

The values of temperature asymmetry are given in Table 6. Corresponding rotational velocities obtaining from Eq.(10) are given in the following Tables 7 and 8.

Even though we know that the dust contribution is highest in 857 GHz frequency band,  we also calculate the velocities for 545 and 353 GHz bands for two different models (DL07 and GNILC). Calculations are performed for constant and non-constant temperature and spectral indexes.

From Tables 7 and 8 one can clearly see, that for DL07 model neither of two different scenarios shows any significant difference, as opposed to GNILC model. For all the models and scenarios we get relatively small values for velocities for 857 GHz, which implies slow or no rotation of galactic dust component.

However, it is possible to roughly estimate the dust velocity in different way. One can assume that the dust component has a Keplerian motion and  that gravity and centrifugal force are in equilibrium, i.e. the orbital velocity is given by
\begin{equation}
V^{2}(r)=\frac{GM}{r},
\end{equation}
where $M$ is the virial mass. 

There are different estimates for dust mass in M31. For example, according to \cite{Haas} the mass is $M_{dust}=3.8\times 10^{7} M_{\odot}$ and in \cite{Kong} $M_{dust}=1.3\times 10^{7} M_{\odot}$ values are given for dust within 18kpc from center of the M31. 

However, for the lower limit of $M_{dust}=1.1\times 10^{7} M_{\odot}$  is derived \cite{Mont} with a best-fit model value of $M_{dust}=7.6\times 10^{7} M_{\odot}$, in agreement with expectations from \textit{CO} and \textit{HI} measurements.
Here we adopt the values given by \cite{Kong} and we obtain velocities in the range of 600-1200 km/s. The derived velocities strongly depend on the frequency bands, the highest values are obtained at 353 GHz, significantly higher than of 857 GHz and 545 GHZ bands. 

\begin{table}[h!]   %%% 7
\tbl{M31 halo rotational velocities according to DL07 model.}
{
\centering
\begin{tabular}{ |p{2cm}||p{2cm}||p{2cm}||p{2cm}|}
\hline
\multicolumn{4}{|c|}{DL07 \(\beta=const\)    T=const} \\
\hline
Distance\newline r (kpc)
&$\nu=857GHz$\newline  $V_{rot}$(km/s)
&$\nu=545GHz$\newline  $V_{rot}$(km/s)
&$\nu=353GHz$\newline  $V_{rot}$(km/s)\\
\hline
  40    & 9.45      & 36.21       & 359.41  \\

  60  &12.6    &70.08       & 643.30  \\
 
80& 13.02        & 99.38      & 474.15  \\
 
100&9.47       &142.41     &  640.54  \\
\hline
\hline
\multicolumn{4}{|c|}{DL07 $\beta=const$    $T\neq const$} \\
\hline
Distance\newline r (kpc)
&$\nu=857GHz$\newline  $V_{rot}$(km/s)
&$\nu=545GHz$\newline  $V_{rot}$(km/s)
&$\nu=353GHz$\newline  $V_{rot}$(km/s)\\ 
  \hline
40    &   9.43    &  36.27      & 357.05  \\

  60  & 12.61   & 70.05      &644.78   \\
 
80&      13.0    & 99.52     & 479.12  \\
 
100&       9.51&   142.35  & 635.2   \\
  \hline
  \end{tabular} \label{table:7}}

\end{table}

\begin{table}[h!]   %%% 8
\tbl{M31 halo rotational velocities according to GNILC model.}
{
\centering
\begin{tabular}{ |p{2cm}||p{2cm}||p{2cm}||p{2cm}|}
 \hline
  \multicolumn{4}{|c|}{GNILC $\beta\neq const$,    $T\neq const$} \\
 \hline
Distance\newline r (kpc)
&\(\nu=857GHz\)\newline  \(V_{rot}\)(km/s)
&\(\nu=545GHz\)\newline  \(V_{rot}\)(km/s)
&\(\nu=353GHz\)\newline  \(V_{rot}\)(km/s)\\ 
  \hline
 40    &   0.34    &  1.85      & 25.06  \\

  60  & 0.51  & 3.55      &50.63   \\
 
80&      0.50    & 7.41     & 84.71  \\
 
100&       0.41&   7.79  & 78.55   \\
  \hline
\hline
  \multicolumn{4}{|c|}{GNILC $\beta=2$,    $T= const$} \\
\hline
Distance\newline r (kpc)
&$\nu=857GHz$\newline  $V_{rot}$(km/s)
&$\nu=545GHz$\newline  $V_{rot}$(km/s)
&$\nu=353GHz$)\newline  $V_{rot}$(km/s)\\
  \hline
 40    &   8.97   &  38.74     & 427.08  \\

  60  & 12.0  & 67.66      &563.87   \\
 
80&      11.36    & 134.89     & 511.65  \\
 
100&      9.22&   140.21 & 536.14   \\
\hline
  \end{tabular} \label{table:8}}
\end{table}

\section{Conclusions}

Whence various details of the rotation of disks of spiral galaxies are well studied, the structure and the rotation of galactic dark halos still remain far less established. Here we attempted to study the halo dynamics using M31 galaxy dust emission data vs the microwave data obtained by Planck; the microwave data can trace dark matter distribution at larger scales, e.g. \cite{GK}. We derived the velocity values for the rotation of the dust component using two different models, DL07 and GNLIC. For 857 GHz frequency band we get relatively small velocity values for both models, while for 545 GHz we obtained up to 145 km/s and it is clear that the change of $T_{d}$ does not affect the velocity values.  It is worth to stress the relatively large values for 343 GHz, up to 650 km/s, which is consistent with our estimate of 600-1200 km/s. The dust emission flux densities which we have calculated from CMB maps have lower values at 545 GHz and 353 GHz bands. Since we can express the dust masses through the observed flux densities and since the flux density at 857 GHz is significantly higher, the major part of the dust emission comes at this band. At this band the majority of dust velocities have low values, i.e. implying rather weak rotation. 

The dust component alone certainly cannot entirely cover the entire radiation emission mechanisms in the galactic halos, however the accurate information on the dust dynamics can help to fill the gap between theoretical and measured velocity curves, including those obtained from the Planck microwave temperature maps.

\section{Acknowledgments}

I am thankful to S. Mirzoyan for valuable discussions and the referee for valuable comments. The use of  Planck data in the Legacy Archive for Microwave Background Data Analysis (LAMBDA) and HEALPix package is acknowledged.

\end{document}